\documentstyle[12pt]{article}
\def\xbar{\bar \xi}
\def\zbar{\bar \zeta}

\def\cF{{\cal F}}

\def\cN{{\cal N}}

\newfont{\goth}{eufm10 scaled \magstep1}

\def\a{\alpha}

\def\c{\gamma}\def\C{\Gamma}
\def\d{\delta}
\def\e{\epsilon}

\def\k{\kappa}

\def\s{\sigma}

\def\beq{\begin{equation}}\def\eeq{\end{equation}}
\def\beqa{\begin{eqnarray}}\def\eeqa{\end{eqnarray}}
\def\barr{\begin{array}}\def\earr{\end{array}}

\def\del{\partial}

\def\dt{\tilde{\d}}
\def\xt{\tilde{x}}

\def \ss {{\s \hspace{-7.4pt} \slash}\;}
\def \ys {{y\kern-.5em / \kern.3em}}
\def \one {{1_{4 \times 4}}}
\def\ads{$AdS_5 \times S^5$ \,}

\newtheorem{higgs}{Higgs Config.}
\newtheorem{brane}{Brane Config.}

\let\bm=\bibitem

\def\nn{\nonumber}
\def\bd{\begin{document}}
\def\ed{\end{document}}
\def\ba{\begin{array}}
\def\ea{\end{array}}
\def\bea{\begin{eqnarray}}
\def\eea{\end{eqnarray}}
\def\ft#1#2{{\textstyle{{\scriptstyle #1}\over {\scriptstyle #2}}}}
\def\fft#1#2{{#1 \over #2}}
\newcommand{\be}{\begin{equation}}
\newcommand{\ee}{\end{equation}}
\newcommand{\eq}[1]{(\ref{#1})}
\def\eqs#1#2{(\ref{#1}-\ref{#2})}
\def\det{{\rm det\,}}
\def\tr{{\rm tr}}
\newcommand{\ho}[1]{$\, ^{#1}$}
\newcommand{\hoch}[1]{$\, ^{#1}$}
\def\ra{\rightarrow}
\def\uha{{\hat {\underline{\a}} }}
\def\uhc{{\hat {\underline{\c}} }}

\begin{document}

\hfill{NEIP-98-013}

\hfill{hep-th/9810195}

\vspace{20pt}

\begin{center}

{\Large\bf  D3 Brane(s) in $AdS_5 \times S^5$ and $\cN =4,2,1$ SYM }
\vspace{30pt}

{\large Adel Bilal and Chong-Sun Chu }

\vspace{15pt}

{\small \em Institute of Physics,
University of Neuch\^atel, CH-2000 Neuch\^atel, Switzerland}

\vspace{60pt}

{\bf Abstract}
\end{center}

Recently, Douglas and Taylor proposed to identify the Higgs vev of the
Coulomb branch of the 4 dimensional $\cN=4$ super Yang-Mills theory with
the positions of D3 branes in the \ads string theory. We extend this
identification to more general configurations that preserve less
supersymmetry. We show that a single D3 brane in \ads string theory can
break the spacetime supersymmetry to 1/4 or even 1/8. Such configurations,
interpretated as backgrounds of the SYM, also break the $\cN=4$
superconformal symmetry of SYM to 1/4 or 1/8, giving $\cN=2$ or
$\cN=1$. We discuss the implications
of this correspondence. We also present other BPS D3-brane configurations 
that preserve 1/8 supersymmetry corresponding to $\cN=1$ on the SYM side.

\pagebreak

\section{Introduction}

A  year ago, a remarkable proposal has been put forwarded by
Maldacena \cite{mal} which conjectures that the type IIB string theory on
\ads is dual to the four dimensional $\cN=4$ supersymmetric Yang-Mills theory
(SYM) with gauge group $SU(N)$. 
A precise recipe of relating type IIB 
sugra  in \ads to the SYM on the $AdS$ boundary has been given in
\cite{gubser,witten1} and it was demonstrated that 
this duality is holographic in nature.
This holographic  property has been further studied
in \cite{witten1,sw,barbon1,mans,rey,hull,minic} and in 
particular in the presence of branes in \ads in 
\cite{witten1,green1,chw,kogan,green2,park,DT,vijay,dorey,wu}.

Recently, Douglas and Taylor \cite{DT} proposed that the large $N$ limit
of the $\cN=4$ $SU(N)$
SYM in the Coulomb branch, i.e. with nonzero vacuum expectation values for
the scalar fields, corresponds to branes in the bulk of \ads string theory. In
the large $N$ limit, one can replace $SU(N)$ by $U(N)$ and these authors
considered  the following configuration of Higgs fields
\be \label{DTconfig}
X^m =\pmatrix{x^m & 0 &0 \cr 0&\xt^m &0 \cr 0 & 0& \bf{0} }\ .
\ee
They provided some evidence that $\cN=4$ SYM in the vacuum \eq{DTconfig} does
describe two D3-branes in the $AdS$ bulk.
This proposal  is in accord with the principle of holography.
The D3-branes these authors considered have worldvolumes parallel to the
$AdS$-boundary.  One may wonder what about more general D3-branes
configurations in the $AdS$ bulk? How are they described in the SYM
picture? This paper is a first attempt to answer these questions. 
We propose that the identification of SYM  configurations with
configurations of D3-branes in \ads 
indeed holds true even for non-constant backgrounds.

Like in most other checks of various dualities, the study of BPS
configurations is a valuable tool.  On one hand it allows one to perform
a manageable analysis of the problem, on the other hand it can often tell
us a lot about the duality and provide non-trivial checks. 
In this paper, we construct BPS configurations preserving 1/2, 1/4 and 
1/8 of the relevant supersymmetries,\footnote{
To avoid confusion let us insist that initially we have 32
supersymmetries. 
This is obvious on the IIB side. 
On the SYM side one has an $\cN=4$ Poincar\'e supersymmetry 
(16 supersymmetries), and at the superconformal point also 16 special 
superconformal symmetries, hence also a total of 32. 
The notation $\cN=4$ SYM thus is a bit misleading at the conformal point.
}
both for the $\cN=4$ SYM theory
and for the 10 dimensional \ads string theory. The constructions are 
independent of each other and a priori, there is no relation 
between these BPS configurations.
We then check  the above  generalized form of 
the duality proposal by showing that there is a precise and consistent 
matching of supersymmetry if one identifies the non-constant Higgs values
and YM field strength 
with the positions and the Born-Infeld field strength of the
D3-brane(s) 
in \ads.

The organization of the paper is as follows. First, within the SYM      
framework, we study
the constant Higgs configurations 
proposed by \cite{DT} in more detail. 
These configurations preserve $\cN=4$
conformal supersymmetry in four dimensions (32 susys) or only
$\cN=4$  supersymmetry (16 susys), depending on whether the Higgs 
vevs are all zero or not.  
Then  we present Higgs configurations that preserve 
$\cN=2$ or $\cN =1$  supersymmetry in four dimensions. 
This is achieved by allowing two (or four)  
of the Higgs scalars to depend holomorphically on two of the worldvolume
coordinates (see condition \eq{holo} below).
Finally we present other different ways of getting $\cN=1$. This is achieved
by adding a self-dual YM configuration on top of the
$\cN=2$ Higgs configuration. We show that these new 
configurations break 7/8 of the supersymmetries and give rise to $\cN=1$. 

Section 3 contains an independent analysis of the supersymmetry
preserved by the corresponding D3-brane in \ads. We first recall the
supersymmetry preserving condition for D$p$-branes.  
This condition can
be derived from the $\k$-symmetric formulation of D$p$-branes. 
A flat D3 brane preserves 1/2 of the supersymmetries.
Then we present a novel brane configuration that preserves 1/4  
supersymmetry. This
D3-brane configuration is indeed a direct transcription of the Higgs
configuration in section 2 by interpretating the latter as the position
of the D3-brane in \ads. We show that the resulting D3-brane in fact
preserves the same amount of supersymmetry as is preserved on the gauge
theory side. A precise and consistent 1-1 mapping between the 4
dimensional supersymmetry and the 10 dimensional $AdS$ supersymmetry is
established.
We also show that our results imply that, by allowing for
two D3-brane probes, one can break the spacetime supersymmetry down to
1/8. 
We then discuss that  by allowing a ``self-dual" Born-Infeld  configuration 
on the worldvolume of a single D3-brane, one can also reduce  the residual 
supersymmetry by half and
this {\it single} D3-brane breaks 7/8 of the spacetime supersymmetries.
Such ``self-dual" configurations make sense with Euclidean signature
but 
their Minkowski interpretation is less clear.

Single D3-brane configurations which preserve 1/4  of the spacetime
supersymmetry were first constructed in \cite{ooguri2} by wrapping the
D3 over certain cycles in a Calabi-Yau. Here, we construct new D3-brane
configurations that preserve 1/4 or 1/8 of the spacetime 
supersymmetry.
One of the differences of our construction with theirs is that the 
D3-brane we construct is not wrapped on any cycle
\cite{bb,ooguri1,ooguri2} 
but infinitely extended.  
For the 1/4 supersymmetries case, we found that the condition for it to
satisfy the equations of motion is precisely the same as the condition
for it to preserve 1/4 sypersymmetry.  On the SYM side this corresponds
to an extra condition for the gauge theory Higgs configuration which is
not necessary in the classical limit. 
For the 1/8 supersymmetries case, we again find that the brane
configuration is slightly more restrictive then the corresponding classical
YM configurations.
The reason and  significance of this are discussed in section 4, where
we summarize our results and present some directions for
further works. 

\section{BPS Configurations in SYM}

Consider $\cN=4$ SYM living in (3+1) dimensions with coordinates 
$\s^i, i=0,1,2,3$. For our purpose here, it is convenient to think of
the 4 dimensional $\cN=4$ SYM theory as obtained through dimensional
reduction of the 10 dimensional SYM theory and use the 10 dimensional
$\C$ matrix notation. The field content consists of the gauge fields
$\hat A_i$, the six scalars $\hat X_m$, $(m=4,\cdots,9)$ and the 16 fermions
$\hat \Psi$. $\hat \Psi$ is a Majorana-Weyl spinor in 10 dimensions and it
decomposes into four Majorana spinors in 4 dimensions. All fields are
Hermitian and are in the adjoint representation\footnote{
Our convention is to put a hat on the symbols for the adjoint $U(N)$ 
matrices. Below the same symbols without hats will be used for a
certain 
diagonal element of these matrices.
} of the gauge group. 
The theory is invariant under 16 Poincar\'e supersymmetries  $\d$ and 
16 special supersymmetries $\dt$ given by 
\bea
\d \hat X_m &=&i\xbar\C_m \hat \Psi, \label{dX1}\\
\d\hat \Psi &=&( \frac{i}{2} [\hat X_m,\hat X_n]\C^{m n} + 
D_i \hat X_m \C^{im} +
\ft12 \hat F_{ij} \C^{ij} ) \xi, \label{dPsi1}\\
\d \hat A_i &=& \xbar \C_i  \hat\Psi, \label{dA1}
\eea
and
\bea
\dt\hat  X_{m}&=&i\zbar \ss\C_m \hat \Psi, \label{dX2}\\
\dt \hat \Psi &=&( \frac{i}{2} [\hat X_m,\hat X_n]\C^{m n} + 
D_i \hat X_m \C^{i m} +
\ft12 \hat F_{ij} \C^{ij} ) \ss \!\zeta
+2\hat  X_m \C^m \zeta, \label{dPsi2} \\
\dt \hat A_i &=& \zbar \ss \C_i \hat \Psi. \label{dA2}
\eea
The $\C$-matrices are $16\times 16$ dimensional, 
Hermitian and satisfy  the Clifford algebra
\be \label{flatG}
\{\C_M, \C_N \} =2 \eta_{M N},\quad M, N =0,\cdots, 9,
\ee
where the metric is $\eta_{M N}=diag (-,+,\cdots,+)$.

\subsection{$\cN=4$ and $\cN=2$ Configurations}

In this paper, we will only consider diagonal Higgs configurations $\hat X^m$ 
with the non-vanishing
eigenvalues all different from each other. The theory reduces to a
number of non-interacting sectors and one can apply the analysis to each
$U(1)$ sector individually. The analysis becomes harder when one allows
for coinciding  eigenvalues. The corresponding brane picture will
be a set of overlapping D3-branes described by a non-Abelian
Born-Infeld 
action. The later hasn't been fully understood \cite{Tsey}. 

Throughout this paper, we
will restrict ourself to   bosonic configurations ($\hat \Psi =0$), 
and until further notice also to
$\hat F_{ij}=0$.  Let us  consider Higgs configurations that 
are diagonal with only the first diagonal element non-vanishing
\be \label{config}
\hat X^m =\pmatrix{X^m & 0 \cr 0 & \bf{0} }.
\ee
In the following, $X^m$ will denote the non-vanishing diagonal entry 
of the matrix
$\hat X^m$, and similarly $\Psi$ and $F_{ij}$ will denote the
corresponding 
diagonal elements of $\hat \Psi$ and $\hat F_{ij}$.
Consider first the
\begin{higgs}  ($\cN=4$)
\label{config1}
\be
X^m= c^m \quad 
\mbox{where $c^m$ are constants and not all zero} \quad
(m= 4, \cdots, 9).
\ee
\end{higgs}
It is easy to  check that this configuration  breaks all the special
supersymmetries and preserves all the 16 Poincar\'e supersymmetries of SYM.
This is the configuration considered in \cite{DT} which is proposed to
describe a D3-brane sitting parallel to the boundary of \ads. In the
brane picture, as we will check below, such a D3-brane  breaks half of
the 32 \ads supersymmetries and the picture is indeed consistent.

We are interested in having $\cN=2$ supersymmetry in 4 dimensions.
The original $\cN=4$ vector multiplet will split into an $\cN=2$ vector 
multiplet (two real scalars)
and an $\cN=2$ hyper-multiplet (four real scalars). The R-symmetry 
is broken as
\be
SO(6)_R \rightarrow SU(2)_R \times U(1)_R \ .
\ee
This suggests to try the following
configuration,
\begin{higgs}  ($\cN=2$)
\label{config2}
\bea
&X^{4,5,6,7} =  \, \mbox{constants, and} \quad
X^{8,9} =  X^{8,9}(\s_1,\s_2) \quad \mbox{satisfying}\\
&\del_1 X^8 =\pm \del_2 X^9,\quad \del_2 X^8 =\mp \del_1 X^9. \label {holo}
\eea
\end{higgs}
(Obviously, we exclude the trivial case where $X^{8,9}$ both vanish.)
Due to the explicit $\s_i$ dependence in \eq{dPsi2},
it is easy to see that with this Higgs configuration the special 
supersymmetries are generally  all
broken even if all of the $X^{4,5,6,7}$ vanish. So to
preserve 1/4 of the total of $16 + 16$ supersymmetries, we will try 
to find conditions on
$X^{8,9}$ such that 1/2 of the 16 Poincar\'e supersymmetries 
are preserved. The latter act as
\be
\d \Psi = \C^{18} (\del_1 X^8 +\C^{12}\del_2 X^8) M \xi,
\ee
where
\be \label{M}
M= 1+ (\del_1 X^8 + \C^{12} \del_2 X^8)^{-1} (\del_1 X^9 + \C^{12}
\del_2 X^9)\C^{89}.
\ee
Notice that $(\del_1 X^8 + \C^{12} \del_2 X^8)$ is invertible as
long as $\del_1 X^8$ or $\del_2 X^8$ is non-vanishing. Thus in the
present case,  \eq{M} is well defined. 
It is not hard to show that $M$ is a matrix of half rank iff
\be
\del_1 X^8 =\pm \del_2 X^9,\quad \del_2 X^8 =\mp \del_1 X^9,
\ee
i.e. $X^8, X^9$ satisfy the Cauchy-Riemann condition. This condition
also guarantees that $X^{8,9}$ satisfy the equations of motion. In this
case, $M$ is equal to
\be \label{mm}
M= 1 \pm \C^{1289}
\ee
and hence only that half of the Poincar\'e supersymmetries $\xi$ 
which satisfy
\be \label{mmm}
(1 \pm \C^{1289}) \xi =0
\ee
are preserved.
We will see in the next section  
that this Higgs configuration corresponds in the brane picture
to having a D3-brane in \ads whose embedding satisfies the Cauchy-Riemann 
condition \eq{holo}.

Notice that a condition similar to \eq{holo} has also appeared in the
construction of the three-brane solition \cite{hlw} on the M5 brane
worldvolume. There, the M5-brane occupies the $\s^{0,1,2,3,4,5}$ directions
and the transverse scalars are $X^{6,7,8,9,10}$.  It was shown that a
three-brane soliton preserving 1/2 of the M5-brane worldvolume
supersymmetry can be constructed if 
\be \label{howe}
\del_4 X^6 =\pm \del_5 X^{10},\quad \del_5 X^6 =\mp \del_4 X^{10},
\ee
and with the other three scalars unexcited. 
The three-brane soliton has worldvolume in the 0123 directions and
the condition is imposed entirely  in the {\it transverse} direction. 
This is different from our case.

The most general configuration with two Higgs scalars excited is

{\bf Higgs Config. 2'} 
{\it 
\label{config3}
$X^{8,9} =  X^{8,9}(\s_0,\s_1,\s_2,\s_3) \,$  
and $\, X^{4,5,6,7} =$ constants.
}

It is possible to show that this configuration satisfies the equations
of motion and preserves half of the Poincar\'e supersymmetries iff it reduces
to the Higgs configuration \ref{config2} and hence the same holomorphicity
condition \eq{holo} applies.

\subsection{$\cN=1$ Configurations}
From the discussion above, it should be clear that 
one way to get  $\cN=1$   is to consider a Higgs configuration of the form
\begin{higgs}  ($\cN=1$)
\be
\hat X^m =\pmatrix{X^m & 0 &0 \cr 0&\tilde{X}^m &0 \cr 0 & 0& \bf{0} }
\ee
with 
\bea
&&X^{8,9}(\s_1,\s_2) \quad \mbox{satisfying the Cauchy-Riemann condition} 
\label{n1}\\
&&\tilde X^{6,7}(\s_1,\s_2) \quad 
\mbox{satisfying the Cauchy-Riemann condition}
\label{n2}\\
&& \tilde X^{8,9}, X^{6,7}, X^{4,5}, \tilde X^{4,5} = \mbox{constants}
\eea
\end{higgs}
The condition \eq{n1} on $X^{8,9}$ gives rise to the condition
\be \label{proj1}
(1 \pm \C_{1289}) \xi =0
\ee
for the unbroken supersymmetries $\xi$. Similarly, \eq{n2} gives rise to
\be \label{proj2}
(1 \pm \C_{1267}) \xi =0
\ee
and the two projectors in \eq{proj1} and \eq{proj2} commute.
The two conditions together break the Poincar\'e supersymmetry to 1/4 and
only a $\cN=1$ supersymmetry is left unbroken.

There are also other Higgs configurations with $\cN=1$ by 
trivially permuting the coordinates $\s_i$ and/or the Higgs fields $X^m$ as
long as we obtain commuting projectors as in \eq{proj1}, \eq{proj2}. 
For example, a configuration with 
$X^{8,9}(\s_1,\s_2)$
and $\tilde X^{6,7}(\s_1,\s_3)$  satisfying Cauchy-Riemann condition (all the 
other $X$, $\tilde X$'s being constant)
wouldn't work and will break all supersymmetry
as  $\C_{1289}$ and $\C_{1367}$ anticommute 
rather than commute with each other.
The same configuration with $X^7$ replaced by $X^9$ has $\cN=1$ 
unbroken supersymmetry as  $\C_{1289}$ and $\C_{1369}$ 
commute with each other.

We will see that in the brane picture, the Higgs configuration 3
corresponds to having two D3-branes in \ads, each satisfying a holomorphic
embedding condition.  

It is  known that  an (anti-)instanton background breaks half of  the special 
supersymmetries and breaks half of the Poincar\'e supersymmetries. 
In this paper, we are  working with Lorentzian signature, 
so we should look at the configuration that is obtained from 
the  Euclidean instanton after continuation to Minkowski space, namely
\be \label{yminst}
\hat{F}_{ij} = \frac{c}{2} \e_{ijkl} \hat{F}^{kl}\, ,\ \ c=\pm i\ .
\ee
The case $c=-i$ corresponds to a (Euclidean)
YM instanton and $c=i$ corresponds to a (Euclidean)  anti-instanton.
We will simply refer to these  ``self-dual" or ``anti-self-dual" 
Minkowskian configurations as (Minkowskian) instantons\footnote{
It is obvious that the Minkowski  (anti) self-duality equation \eq{yminst}
leads to certain components of $F_{ij}$ that are purely 
imaginary. The proper way to think about these  configurations 
is that they make sense when continued to Euclidean signature. 
However, since throughout this paper we work with Minkowski 
signature, we will formally use \eq{yminst}.
}
Let us first embed an instanton in the non-abelian 
part of $\hat{F}$ and consider the following configuration

{\bf Higgs-Instanton Config. 4} ($\cN=1$)
{\it
\be 
\hat X^m =\pmatrix{X^m & 0 \cr 0 & \bf{0} },
\quad \hat F_{ij} = \pmatrix{0 & 0 \cr 0 & F_{ij} }
\ee
where $X^m$ is the Higgs configuration  2 above and
$F_{ij}$ is an (anti-)instanton. 
}

As we will see shortly, the unbroken supersymmetry is $\cN=1$.

Next, we would like to consider a configuration where the non-trivial 
gauge field lies in the same $U(1)$ factor as the non-trivial Higgs 
component. It is well-known that one cannot embed an 
instanton into the $U(1)$ factors of a gauge 
group. However, one can consider the following (anti) self-dual configuration
\bea \label{asd}
&\hat F_{ij} = \pmatrix{F_{ij} & 0 \cr 0 & \bf{0} }, \\
&F_{ij} = \frac{c}{2} \e_{ijkl} F^{kl}\, , \ c=\pm i
\eea

and consider the following

{\bf Higgs-Gauge Config. 5} ($\cN=1$)
{\it
\be 
\hat X^m =\pmatrix{X^m & 0 \cr 0 & \bf{0} },
\quad \hat F_{ij} = \pmatrix{F_{ij} & 0 \cr 0 & \bf{0} }
\ee
where $X^m$ is the Higgs configuration  2 above and
$F_{ij}$ is the self-dual or anti self-dual configuration \eq{asd}
}

For both configurations 4 and 5 above,
one easily checks that the equations of motion are satisfied and  the special
supersymmetries are all broken. From eq. \eq{dPsi1} we see that the preserved 
Poincar\'e supersymmetries must satisfy
\bea 
&&(1+ c \C_{0123}) \xi =0\\
&&(1 \pm \C_{1289})\xi =0, 
\eea
and hence there is an unbroken $\cN=1$  supersymmetry left. 
The corresponding brane picture will be discussed in the next section.
In particular, we will see there that the self-dual configuration 
5 corresponds to a
single D3-brane with a  self-dual Born-Infeld field strength.
Note that such  constructions with a non-vanishing $\hat F$  field could 
also be done for the Higgs configuration 1 resulting in a $\cN=2$ SYM.
The case $X^m=0$ was the subject of interest in \cite{green1}.

\section{D3-Branes in \ads}

D-branes are basic objects in the nonperturbative formulation of string
theory. They are endpoints of open strings and one can learn much of
their properties by studying the boundary (S)CFT of the open strings
ending on it. Along this line, \cite{ooguri1,ooguri2}
used the SCF worldsheet description to derive the embedding condition
for a D-brane to preserve supersymmetry. There has been some progress
\cite{adss1,adss2bis,adss2,adss3} in constructing \ads string theory. However a
satisfactory string worldsheet description including non-trivial RR
backgrounds is still lacking. Therefore although one would like to carry
out a microscopic SCFT analysis similar to \cite{ooguri1,ooguri2} to see
how a D3-brane in \ads breaks the spacetime supersymmetry, we will be
contented in this paper with a low energy analysis,  using a
$\k$-symmetric formulation of D-branes. 

We will present configurations of a single D3-brane in \ads 
which preserve 1/2, 1/4 or 1/8 of the spacetime
supersymmetries. Until now, the only know way for a single D-brane to
break more than half of the spacetime supersymmetries was to wrap the
brane over a Cayley submanifold \cite{ooguri2}. As these authors have
shown, such a brane saturates a BPS condition and hence attains its
minimal energy. For \ads, there is no appropriate cycle on
which the D3-brane could be wrapped. So our construction has to be
different. In fact, we find that the desired configuration 
(\eq{phys}-\eq{x89} below)  preserving 1/4 of the spacetime supersymmetries is
non-compact. Since the brane is not wrapped on anything, one also needs to
make sure it satisfies the equations of motion. 
Maybe somewhat surprisingly, we find
that the condition for it to satisfy the equations of motion is precisely
the same condition for it to preserve  supersymmetry. The
configuration \eq{phys}-\eq{x89} provides the only known example of a
single, unwrapped D3-brane preserving 1/4 of the spacetime
supersymmetries. By turning on an appropriate ``self-dual" 
Born-Infeld field strength $\cF$ on a single D3-brane, 
we will also get a configuration preserving 1/8 of the supersymmetries.

\subsection{Supersymmetry Preserving Condition}

\underline{$\kappa$-symmetry on a D$p$-brane}

In order to be self contained, we briefly review here the necessary part
of the $\k$-symmetric formulation for a D$p$-brane coupled to a
supersymmetric background. For details of the formulation as an
action, see \cite{d0,d1,d3,d4,d2}; 
and for the
formulation as equations of motion, see \cite{hs,hrss}. 
For a pedagogical review we refer to \cite{sezgin}, while
the equivalence of these approaches is discussed in \cite{hrs}. 
We will not need the details here. What we need is that
the $\k$-symmetry  takes the simple covariant form 
\be \label{kappa}
\d \theta = (1+\C)\k,
\ee
where $\theta$ is the spacetime spinor depending on the worldvolume
coordinates $\s$, $\k(\s)$ is a local parameter for the
$\k$-transformation and $\C$ is the ``pull-back'' $\C$-matrix which
depends on the worldvolume fields and satisfies $\C^2=1$ (see below). 

For our present use, we recall the explicit form
of $\C$ \cite{bkog} in the case that all fermionic fields vanish.
It is
\be \label{aga}
\C = e^{-a/2} \C_{(0)}' e^{a/2},
\ee
where $a$ is a matrix given below and
\be \label{gamma0'}
\C'_{(0)}=\cases{ (\C_{11})^{p-2\over 2} \C_{(0)}& IIA,\cr
(\s_3)^{p-3\over 2} i \s_2\otimes \C_{(0)}\ &IIB,}
\ee
and
\be \label{gamma0}
\C_{(0)}= \frac{1}{(p+1)! \sqrt{|g|}} \epsilon^{i_1\cdots i_{(p+1)}}
\del_{i_1} X^{M_1} \cdots \del_{i_{(p+1)}}X^{M_{(p+1)}}
\C'_{M_1 \cdots M_{(p+1)}}.
\ee
As usual, the matrix  $\C'_{M_1 \cdots M_{(p+1)}}$ is the 
antisymmetrized product of the $\C_{M_k}'$ with the $\C_{M}'$ 
being the 10 dimensional $\C$-matrices in the
coordinate basis defined by
\be
\C'_M := E_M{}^A \C_A,
\ee
where the $\C_A$ are flat space $\C$-matrices.
The metric $g_{ij}$ is the induced worldvolume metric
\be
g_{ij} = \del_i X^M \del_j X^N G_{MN}
\ee
and $|g|$ is its determinant.
To define the matrix $a$ appearing in \eq{aga} we need to introduce  
the modified 2-form field strength $\cF$ which is related to the Born-Infeld
field strength $F= dA$ by
\be
\cF =F - \underline{B},
\ee
where $\underline{B}$ is the pull-back of the target space NS-NS
2-form potential to the worldvolume.
The matrix $a$ depends only on the worldvolume Born-Infeld field
strength and is given by
\be \label{a}
a = \cases{-{1\over2} Y_{jk} \c^{jk} \C_{11} & IIA, \cr
	    {1\over2} Y_{jk} \s_3\otimes\c^{jk}&IIB,}
\ee
the $\c^{jk}$ being worldvolume $\c$ matrices, 
\be
\c_i = \del_i X^M \C'_M
\ee 
and  $Y$ is a function of $\cF$. The relation in the frame basis of 
the worldvolume (underlined indices) is 
\be
Y_{\underline{j}\underline{k}}:= 
\mbox{tan}^{-1} \cF_{\underline{j}\underline{k}}.
\ee
One can show that $\C$ satisfies,
\be
\mbox{tr}\, \C=0,\quad \C^2=1.
\ee

We will be interested in the case of a D3-brane in a IIB background.
A general D3-brane configuration will break  part
or all of the supersymmetries of the  10 dimensional background.
The surviving supersymmetries must  satisfies \cite{bkog} 
\be \label{cond1}
(1-\C) \xi =0
\ee
since only then can the $\kappa$-symmetry \eq{kappa} compensate 
for the transformation induced by $\xi$, i.e. $\xi=-(1+\C)\kappa\, $ 
for some $\kappa$.
The existence and the amount of unbroken supersymmetry 
impose severe conditions on the brane configuration. 
In this paper, the D3-brane is treated  as a brane probe in the
\ads background (this is appropriate in the limit of large $N$) and 
so $\xi$ are
just the 32 supersymmetries of this background. 

\underline{The \ads background}

The \ads  background of IIB relevant to Maldacena's proposal
\cite{mal} is given
by the metric of \ads and a nontrivial RR 5-form field strength.
Denote the 10 dimensional coordinates by $X^0,\cdots, X^9$. 
The metric is 
\be \label{metric} 
\frac{1}{R^2} ds^2 =	\frac{1}{V^2} (-(dX^0)^2 +\cdots + (dX^3)^2)
	+ \frac{1}{V^2}(dV^2 + V^2 d\Omega_5^2), 
\ee
where 
\be \label{V}
V^2 = (X^4)^2 +\cdots +(X^9)^2
\ee
and the $AdS$-radius is
$R/\sqrt{\a'}= (4 \pi g_{YM}^2 N)^{1/4}$.  The $AdS_5$ is described by
the coordinates $X^{0,1,2,3}$ and $V$ while the $S^5$ is parameterized
by the five
angles of $\Omega_5$.  The non-vanishing RR 5-form field strength is 
\be \label{F5}
F_5 = \frac{4}{R} (\e_{AdS_5} + \e_{S^5}),
\ee
where $\e_{AdS_5}$ and $\e_{S^5}$ are the volume forms on the $AdS_5$ and
$S^5$ respectively. Explicitly, for example
\be \label{F0123V}
F_{0123V} = \frac{4R^4}{V^5}.
\ee
For convenience,  we will adopt a unit of $R=1$ from now on. Explicit
factors of $R$ can be put back easily by simple dimensional arguments.
This background can be thought of as the near horizon limit of the sugra
solution of a D3-brane. See for example \cite{stelle} for a  review
on BPS branes in supergravity. 

The gravitino $\psi_M$ (complex-Weyl) transforms as
\be
\d \psi_M = D_M \xi + \frac{i}{480} \C^{'M_1 \cdots M_5} F_{M_1 \cdots
M_5} \C'_M \xi,
\ee
where $\xi$ is a 32 component complex spinor of positive chirality,
\be \label{chiral11}
(1-\C^{11}) \xi =0, \quad \C^{11} := \C^{01\cdots 9}.
\ee
One can take the following representation for the 
10 dimensional $\C$-matrices which is adapted to \ads,
\bea \label{repr}
\C_\mu &=& \sigma_1 \otimes \one \otimes \c_\mu, \quad 
\mu =0,1, \cdots, 4, \\
\C_m &=& \sigma_2 \otimes \c_m \otimes \one, \quad 
m =5,6, \cdots, 9, 
\eea
where $\s_i$ are Pauli matrices and $\c_\mu$ ($\c_m$) are the 5
dimensional Dirac matrices satisfying
\be
\{\c_\mu, \c_\nu \} = 2 \eta_{\mu \nu}, \quad 
\{\c_m, \c_n \} = 2 \d_{mn}, 
\ee
with $\c_4 := -i \c_{0123}$, $\c_9:= \c_{5678}$.

In this representation,
\be
\C^{01\cdots 9}= \s_3 \otimes \one \otimes \one,
\ee
and \eq{chiral11} is solved by spinors of the form
\be
\xi =  \pmatrix{1 \cr 0} \otimes \e \otimes \eta
\ee
with $\e$ and $\eta$ being  arbitrary $SO(1,4)$ and $SO(5)$ spinors.
The IIB background has 32 supersymmetries of the form,
\bea 
\xi_1 &=&   \pmatrix{1 \cr 0} \otimes \frac{1}{\sqrt{V}}
\e_0^+ \otimes \eta, \label{xi1}\\
\xi_2 &=&  \pmatrix{1 \cr 0} \otimes
(\sqrt{V} + \frac{1}{\sqrt{V}}\ss) \e_0^- \otimes  \eta, \label{xi2}
\eea
where $\e_0^{\pm}$ are 4 components constants spinors satisfying
\be \label{0123}
-i \c_{0123} \e_0^{\pm} =\pm \e_0^{\pm},
\ee
so that $\e_0^+ /\sqrt{V}$ and $(\sqrt{V} + \ss /\sqrt{V})
\e_0^- $ are the Killing spinors of $AdS_5$, 
and $\eta$ is the Killing spinor of $S^5$. It has the form
\be
\eta = \Omega \eta_0,
\ee
where $\eta_0$ is an arbitrary  constant spinor and $\Omega$ is a
certain rotational matrix whose details can be found in \cite{pope}.

The question of preserved supersymmetries then simply 
boils down to finding out which of the $\xi_1$ and $\xi_2$ satisfy 
eq. \eq{cond1}, namely $\C\xi=\xi$.

\subsection{1/2 and 1/4 BPS Brane Configurations} 

We are now ready to analyze the supersymmetry preserved by a bosonic 
D3-brane in \ads. Note that the NS-NS two-form $B$ 
vanishes for our background. 

We will consider $F_{ij}=0$ as suggested by the SYM analysis above and hence 
\be
\cF_{ij} =0
\ee
so that the matrix $a$ vanishes.
It is convenient to work with the complex spinor formalism in
which the factor  of Pauli matrices $\s_i$, for example in 
\eq{gamma0'} and \eq{a}, can be avoided. 
For a spinor $\psi=\pmatrix{\alpha\cr\beta\cr}$ the complex spinor is
$\psi_c=\alpha+i\beta$, and the correspondence is:
$i\s_2 \psi \leftrightarrow -i \psi_c$, 
$\s_3 \psi \leftrightarrow \psi_c^*$.
Eqs \eq{aga} and \eq{gamma0'} simply become
\be \label{gammaeq}
\C=i\sigma_2\otimes \C_{(0)} \leftrightarrow -i \C_{(0)} \ .
\ee
All the D3 brane configurations we will consider will 
always be in the  physical gauge 
\be \label{wvgauge}
X^i=\sigma^i \, , \ i=0,1,2,3 \ .
\ee
We start with a D3-brane sitting parallel to the $AdS$-boundary,
\begin{brane} (1/2 BPS)
\be
X^m= c^m  \quad 
\mbox{where $c^m$ are constants and not all zero}, \quad
(m= 4, \cdots, 9).
\ee
\end{brane}
It is trivial that the brane wave equation of motion is satisfied.
The $\k$-symmetry $\C$ matrix \eq{gammaeq} reduces in this case to 
\be
\C= -i \C_{0123}.
\ee
Thus by eqs \eq{cond1}, \eq{repr} and \eq{0123} we see that
$\xi_1$ is preserved and $\xi_2$ is projected out. The
corresponding SYM theory is in the  Coulomb branch with 16 preserved
(Poincar\'e) supersymmetries. All the special supersymmetries are broken
by the Higgs vev. Thus, we see the following identification of
supersymmetries,

\hspace{4cm}
\begin{tabular}{ccc}
&&\\
\underline{SYM} & & \underline{$AdS_5 \times S^5$}		 \\
Poincar\'e supersymmetry $\xi$ & $\leftrightarrow	$ & $\xi_1$  \\
Special supersymmetry $\zeta$  & $\leftrightarrow	$ & $\xi_2$
\end{tabular}
\be \label{susymap} 
\ee

The identification of supersymmetry between that
of $\cN=4$ SYM and that of the \ads sugra has been pointed out
previously in \cite{chw}. We will see it in more detail
below. We next consider

{\bf Brane Config. 1$'$} {\it (Unbroken SUSY)}
{\it
\be 
X^m= 0 \quad   (m= 4, \cdots, 9).
\ee
}
This means $V=0$. 
It is interesting to note that if we let the D3-brane  go to the
$AdS$-boundary, $V\rightarrow 0$, then 
\bea \label{check}
(1 +i \C_{0123}) \xi_2 & \sim& 
\frac{1}{\sqrt{V}} (1 +i \c_{0123}) \ss \e_0^- \nn\\
&=& \frac{1}{\sqrt{V}} \ss ( 1 -i \c_{0123}) \e_0^- \nn\\
&=& 0
\eea 
due to \eq{0123}.
Thus $\xi_2$ is no longer projected out and we recover the full spacetime
supersymmetry. This is consistent with the SYM picture of going back to
the superconformal phase by letting the Higgs vev $X^m$ go to zero.
The special supersymmetries are  recovered in this limit. 
Note that the structure of the ``$AdS_5$'' part of the  
Killing spinor $\xi_2$ 
(i.e. $(\sqrt{V} + \frac{1}{\sqrt{V}}\ss) \e_0^- $ ) 
is crucial here for this
recovery of the special supersymmetries. It will not work, for example, for
the Killing spinor of a flat metric or a multi-centered solution.

Next we consider the more general case with the  D3-brane embedded as
\begin{brane} (1/4 BPS)
\label{curved}
\bea
&&X_i = \s_i , \quad i=0,1,2,3, \label{phys} \\
&&X^{4,5,6,7} = 0, \label{xeq0}\\
&&X^8 =X^8(\s_1,\s_2), \quad X^9 =X^9(\s_1,\s_2), \label{x89}
\eea
with
$X^8, X^9$ satisfying the condition \eq{holo}. 
\end{brane}

Let's first start with the slightly more general condition
\be \label{xneq0} 
X^{4,5,6,7} = c^{4,5,6,7} \quad \mbox{arbitrary constants}
\ee
instead of \eq{xeq0}. We will see shortly that we need 
$X^{4,5,6,7}$ to be zero for two different reasons. We need it
for the brane to preserve 1/4 supersymmetry, and we also need it 
for the brane  to satisfy the  equations of
motion. We will comment 
on the significance of this requirement on
the SYM side  in the discussion section.

We first look at the supersymmetry preserving condition. 
Substituting \eq{phys}, \eq{x89} and \eq{xneq0} into the definition of 
$\C_{(0)}$, we get, 
\bea \label{gamma00}
\C_{(0)} 
&=& \frac{1}{4! \sqrt{|g|}} \epsilon^{i_1\cdots i_4}
\del_{i_1} X^{M_1} \cdots \del_{i_4}X^{M_4} \C'_{M_1 \cdots M_4} \\
&=&\frac{1}{V^4 \sqrt{|g|}} (Q_0 + \a_1 Q_1 +\a_2 Q_2 + \a^2 Q_3),
\eea
where
\bea
&\a_i := \del_i X^8, \quad i=1,2,\\
&\a^2 :=\a_1^2+\a_2^2, 
\eea
and
\bea
& g_{00} =g_{33} = \frac{1}{V^2}, \quad
g_{11}= g_{22} = \frac{1}{V^2} (1+\a^2), \\
&V^4 \sqrt{|g|} = 1+ \a^2.
\eea
The $Q's$ are numerical matrices
\bea
&Q_0 :=\C_{0123},\quad Q_3 :=\C_{0389},\\
&Q_1 :=\C_{0183} \mp \C_{0923}, \quad Q_2 :=\C_{0823} \pm \C_{0193}.
\eea
(where the $\pm$ signs are correlated with the 
choice of sign in the condition \eq{holo}) and satisfy
\bea
&Q_0^2=Q_3^2= -1, \quad Q_1^2 =Q_2^2 =-2(1 \pm Q_4), \label{Q1}\\
&Q_0 Q_1 + Q_1 Q_0  = Q_0 Q_2 + Q_2 Q_0  =0, \label{Q2}\\
&Q_1 Q_2 +Q_2 Q_1 =Q_1 Q_3 +Q_3 Q_1 =Q_2 Q_3 +Q_3 Q_2 =0, \label{Q3}\\
&Q_0 Q_3 =Q_3 Q_0 = Q_4, \label{Q4}
\eea
with
\be
Q_4 :=\C_{1289}. 
\ee
With this $\C_{(0)}$, and remembering that $\C =-i \C_{(0)}$,
it is not hard to see that $\xi_2$ cannot satisfy the 
condition for unbroken supersymmetry \eq{cond1} and hence the corresponding 
supersymmetries are all
broken. Notice that $\xi_1$ satisfies
\be
(1 + i \C_{0123}) \xi_1 =0 \ \Leftrightarrow \ \xi_1= -i Q_0 \xi_1
\ee
because of \eq{0123} and thus the condition  \eq{cond1} reads
\be \label{immed}
-i (\a_1 Q_1 + \a_2 Q_2) \xi_1 = \a^2 (1+ i Q_3) \xi_1.
\ee
Apply $(\a_1 Q_1+ \a_2 Q_2)$ to the l.h.s. of \eq{immed} to get
\be
(\a_1 Q_1 +\a_2 Q_2)^2 \xi_1 =0,
\ee
where the facts that $Q_3$ anticommutes with $Q_1, Q_2$ and $Q_3^2=-1$ 
have been used. Using \eq{Q1} and \eq{Q3}, we have 
\be \label{1289}
(1 \pm Q_4)\xi_1 =0.
\ee
One can check that this  is also the sufficient 
condition for $\xi_1$ to satisfy \eq{cond1}. Note that $Q_4$ is a
constant matrix, while $\xi_1$ is a spinor that involves the Killing
spinor $\eta$ of $S^5$ at the point on $S^5$ were the D3 brane
intersects it. Thus, in general, it depends nontrivially on the values
of $X^{4,5,6,7,8,9}$, and in particular on the worldvolume coordinates
$\sigma^{1,2}$.  This dependence comes in through the factor ${1\over
\sqrt{V}}$, but more importantly through the matrix $\Omega$ that
relates the $S^5$ Killing spinor $\eta$ to the constant spinor $\eta_0$.
It is thus clear that in general, \eq{1289} will not have any solution
and there will be no unbroken supersymmetry, unless $(1 \pm Q_4)$
commutes with this $\Omega$ matrix. 
This will be the case only if 
\be \label{4zero}
X^{4,5,6,7} =0,
\ee
since then  $\Omega$ is reduced to a simple rotational matrix in the
$X^8-X^9$ plane and hence commutes with $\C_{1289}$. So if one imposes also
the condition 
\be \label{flat2}
(1 \pm \C_{1289}) \e_0^+ \otimes \eta_0 =0
\ee
on the constant spinor $ \e_0^+ \otimes \eta_0$, then we can solve
\eq{1289} for half of the $\xi_1$ and hence the brane configuration 2
preserves 1/4 of the supersymmetries. Notice that the two conditions satisfied
by $\e_0^+ \otimes \eta_0$,
\be \label{two}
(1 + i \C_{0123}) \e_0^+ \otimes \eta_0 =0,
\quad  (1 \pm \C_{1289}) \e_0^+ \otimes \eta_0 =0
\ee
are similar to the usual 1/4 supersymmetry conditions \cite{douglas}
which appear for orthogonally intersecting D-branes in flat target
spacetime. Here however, the background is curved and we have 
a single D3-brane.

Note  that the two conditions 
\be \label{two'} 
(1 + i \C_{0123}) \xi =0,
\quad  (1 \pm \C_{1289}) \xi =0
\ee
satisfied by the unbroken supersymmetries $\xi$ 
are precisely the same as the supersymmetry
preserving conditions in the four-dimensional SYM.
The first eq. \eq{two'} says that $\xi_1$ is kept and $\xi_2$ is projected out.
Using the mapping \eq{susymap},
this is translated to the condition that only the Poincar\'e supersymmetry
$\xi$ can be unbroken. The second eq. \eq{two'} says that $\xi_1$ is
annihilated by $ (1 \pm \C_{1289})$, which is precisely the same
condition \eq{mmm} on the SYM side.
Thus we see that there is a precise matching of unbroken
supersymmetries between certain backgrounds in $\cN=4$ SYM and certain
configurations of a (single) D3-brane in \ads. Not just the amount of unbroken
supersymmetries is equal, but  even the part which is preserved can be
identified consistently.

Finally we check that the brane configuration 2 satisfies the equations of
motion.
Naively, one might expect a simple Laplace equation in \ads  for the
fields $X^n$ (suitably
amended with terms to describe the interaction with the RR-background). 
This is suitable for describing a scalar propagating in
a fixed geometry but is not suitable for the $X^n$, which are themselves
coordinates of the target space.  
In our case, since we are using the physical gauge
$X^{0,1,2,3}=\sigma^{0,1,2,3}$,
the correct equations of motion are
derived from the gauge fixed form of the Dirac-Born-Infeld 
action including the WZ-term, 
\be \label{dbi}
I = \int d^4 \s \sqrt{-{\rm det} (g_{ij} + \cF_{ij}) } + \int \underline{C} 
\; := I_{DBI} + I_{WZ}, 
\ee
where 
\be
g_{ij} = G_{ij} + G_{mn} \del_i X^m \del_j X^n
\ee
is the induced worldvolume metric in the physical gauge and
$\underline{C}$ is the pullback to the worldvolume of the RR
4-form potential. The field strength of $C$ is given by \eq{F5}.
The equations of motion obtained from \eq{dbi} are 
nonlinear in the field $X^n$ and are much 
more complicated than a simple linear Laplace equation. 
We don't know of a general way of solving them but 
we can check whether they are satisfied. 
Define for convenience 
\be
I_{;n} := \d I/\d X^n - \del_j (\d I/\d (\del_j X^n)).
\ee
For configurations satisfying \eq{x89}, we  simply get
\bea
&&I_{DBI;n} = -\frac{4 X^n}{V^6} (1+\a^2), \quad n=4,5,6,7, \\
&&I_{DBI;n} = -\frac{4 X^n}{V^6}, \hspace{2cm} n=8,9, 
\eea
and  for $n=4,\cdots,9$
\bea
I_{WZ;n} &=&
\frac{\del X^{m_0}}{\del \s_0} \cdots \frac{\del X^{m_3}}{\del \s_3} 
 F_{n m_0 \cdots m_3} \nn\\
&=& \frac{X^n}{V} F_{V 0123} \nn\\
&=& \frac{4 X^n}{V^6}
\eea
where eqs \eq{F5} and \eq{F0123V} have been used.
Hence 
\bea
&&I_{;n} = -\frac{4 \a^2 X^n}{V^6}, \quad n=4,5,6,7, \\
&&I_{;n} = 0, \hspace{2cm}  n=8,9 .
\eea
The Euler-Lagrange equations $I_{;n}=0$ for all $n$ thus demand that
$X^{4,5,6,7} =0$, exactly the same condition as we 
got for preserving 1/4 of the supersymmetries.

Before we move on, 
it is important to stress that in performing the checks of this
paper, we have made the identification \eq{V}, instead of the one
\be \label{U}
U^2 =(X^4)^2 +\cdots +(X^9)^2.
\ee
used in \cite{DT}. Their $U$ is our $1/V$. For example, consider the brane
configurations 1 and 1' of a D3-brane sitting parallel to the boundary.
Had we used \eq{U}, then the conformal point ($X^m=0$) of SYM will be
identified with $U=0$, (i.e. $V=\infty$ in our notation). Then in this limit
$\xi_2$ would be $\sqrt{V} \e_0^-$ 
(rather than our ${1\over \sqrt{V}} \ss \e_0^-$)
and the whole $\xi_2$ would be projected out  in the limit.
Our identification \eq{V} was also used in \cite{green2}  
to  reproduce the YM instanton measure from the
D-instanton action in \ads. 

\subsection{1/8 BPS Brane Configurations}

From the  discussion above,  
it should be obvious that 
the following configuration  is 1/8 BPS,
\begin{brane} (1/8 BPS)

Two D3-brane probes in \ads, one 
satisfying the conditions \eq{phys}-\eq{x89} and the other satisfying 
the analogous conditions with the roles of $X^{8,9}$ and $X^{6,7}$ 
interchanged.
\end{brane}
The reason is obvious, the {\it two} D3-branes give 
rise to the following conditions respectively
\bea 
&&(1+ i \C_{0123}) \e_0^+ \otimes \eta_0 =0, 
\quad (1\pm\C_{1289}) \e_0^+ \otimes \eta_0=0,  \label{three1}\\
&&(1+ i \C_{0123}) \e_0^+ \otimes \eta_0 =0, 
\quad (1\pm\C_{1267}) \e_0^+ \otimes \eta_0=0. \label{three2}
\eea
As a result, the brane system preserves 1/8 of the spacetime supersymmetries. 
The three conditions \eq{three1}, \eq{three2} on the constant
spinor  $\e_0^+ \otimes \eta_0$  are precisely the same as the conditions
one would get for {\it three}  D3-branes in flat space-time, and
with worldvolumes in the 
0123, 0389 and 0367 directions respectively. 

It is instructive to recall that in the case of
flat target spacetime, one can break the spacetime supersymmetry to 1/8
by having a triple intersection of D3-branes. Here, we produced the
mathematically similar structure \eq{three1}, \eq{three2} by having two
D3 branes in \ads.  
This suggests to look at other ways to get the same amount of 
supersymmetry but with a D3-brane and another D$p$-brane, 
for example a D-instanton. The desired combination of a D-instanton 
and a D3-brane ({\bf Brane Config. 4}) can be accounted for by 
the Higgs-Instanton configuration 4 of the last section. The 
D-instanton corresponds to the non-abelian YM instanton 
configuration \cite{green1}, while the D3-brane corresponds to the 
Higgs configuration.

As suggested by the Higgs-Gauge configuration 5, 
we now discuss a non-vanishing (abelian) Born-Infeld field strength 
on the D3-brane. In particular, we consider the  ``(anti) self-dual" 
Born-Infeld field $\cF$ satisfying
\footnote{Notice that for the IIB background we are interested in, $B$ is 
zero and so  the Born-Infeld field strength is $\cF=dA$.
}
\be \label{bii}
\cF_{ij} = \frac{c}{2} \e_{ijkl} \cF^{kl},
\ee
with  $c=-i$ for the self-dual and $c=i$ for the anti self-dual case.
This is the simplest and most natural definition one can adopt 
in a nonlinear theory like the Born-Infeld one. 
The same remarks concerning the Euclidean continuation as in 
the last section apply. 

\vspace{.1cm}
Now consider the following {\it single} D3-brane with

{\bf Brane Config. 5} {\it (1/8 BPS)}

{\it
The positions $X^m$ of the D3-brane  satisfy the 
conditions of brane configuration 2 
and the Born-Infeld $\cF$ is self-dual: \eq{bii} with $c=-i$.
}

In this case, $\C$ is given by
\be
\C = e^{-\s_3 \otimes \Delta} (\s_2 \otimes \C_{(0)} ) 
e^{\s_3 \otimes \Delta} 
\ee
where
\be
\Delta :=   Z \cdot  (1-c \C_{(0)})
\ee 
where 
$\C_{(0)}$ is given by \eq{gamma00} and
$Z$ is some matrix depending on the Born-Infeld field strength whose
details are not important to us.
Remember that
\be
\s_3 \otimes \Delta \xi \leftrightarrow \Delta \xi^*
\ee
So if we impose the reality condition on $\xi$
\be
\xi = \xi^*
\ee
and impose 
\be \label{c1}
(1-c \C_{(0)}) \xi =0
\ee
then  the supersymmetry preserving condition
\be \label{c2}
(1-\C) \xi=0
\ee
becomes
\be \label{c3}
(1+ i \C_{(0)}) \xi =0.
\ee 
Equations \eq{c1} and \eq{c3} are only compatible for $c=-i$, 
so only then can one have any unbroken 
supersymmetry (1/8 of the original 32). 
We will comment on this in the discussion section.
Note that such a construction with a non-vanishing $\cF$ field 
could also be done for the brane configurations 1 and 1' 
resulting in 1/4 or 1/2 preserved supersymmetries.

Summarizing, we propose the following 
duality map between {\it backgrounds} in SYM
and brane configurations in \ads:

\hspace{2cm}
\begin{tabular}{lcl}
&&\\
\underline{$SYM$} & & \underline{$AdS_5 \times S^5$}		 \\
Higgs scalars $X^m$ & $\leftrightarrow $ & position $X^m$ of D3-branes\\
YM field strengths $F_{ij}$ &$\leftrightarrow$ & Born-Infeld 
strength $\cF_{ij}$ on the D3-branes
\end{tabular}
\be \label{branemap} 
\ee

\section{Discussion}

In this paper, we presented D3-brane configurations that preserve
$1/n$ (with $n = 2,4,8$) of the original 32 spacetime
supersymmetries. These are in perfect correspondence with the SYM
Higgs/Gauge configurations that preserve  $\cN=4,2,1$ supersymmetry of
the original $\cN=4$ {\it conformal} supersymmetry.  We demonstrated
that when one identifies the values of the Higgs scalars and the
YM field strength $F_{ij}$ with the
positions and the Born-Infeld field strength $\cF_{ij}$ 
of the D3-branes in \ads, one can match not just the amount
of unbroken supersymmetry, but also establish consistently a 1-1
mapping between the 4 dimensional Poincar\'e and special
supersymmetries on the one side, and the spacetime supersymmetries of
\ads on the other side.  In particular, for example,  
we have shown that a single
unwrapped D3-brane in \ads satisfying the Cauchy-Riemann condition
\eq{x89} breaks 3/4 of the spacetime supersymmetries. 

Given the nice matching of supersymmetries for the examples studied
here, it is natural to think that this matching and the proposal of
\cite{DT} will work for the more general cases with general
non-vanishing YM and Born-Infeld field strengths, and probably even to
full generality with nonvanishing fermion fields.

We saw in the last subsection that while the D3 brane with a self-dual
Born-Infeld configuration preserved 1/8 supersymmetry, the anti
self-dual one breaks all of it. Let us now comment on this.  It has
been shown \cite{vijay} that a fundamental (F) string or a D-string
extended in the $\s_i$ directions can be represented by an electric or
magnetic flux of the Born-Infeld field strength. Hence the self-dual
brane configuration 5 can be identified with a F1 - D1 - D3 system,
while the anti self-dual one can be identified with a F1 - anti-D1 -
D3 system. So one can also understand from this point of view why the
first preserves some supersymmetry, while the latter breaks it all.

Due to its importance, we stress again that the condition \eq{xeq0},
$X^{4,5,6,7}=0$  is crucial for the brane configuration 2 to preserve
supersymmetry and to satisfy the equations of motion. However, on the
SYM side, there seems to be no need to impose this condition in order
to preserve 1/4 supersymmetry.  All we need is the holomorphicity
condition \eq{holo}.  We recall \cite{gubser,witten1} that in the
AdS/CFT correspondence, it is the classical tree level sugra action
that becomes the generating functional for the full quantum gauge
theory correlation functions in the large $N$ limit with $g_{YM}^2 N$
fixed but large.  For example as studied in \cite{witten1,Ferrara1},
the Chern-Simon couplings on the sugra side generate the anomalous
correlation functions for the R-symmetry current of the four
dimensional SYM. Our check of supersymmetry in section 2 is a
classical analysis. This is expected to be modified in a full quantum analysis.
Upon quantizing the SYM theory in our non-trivial background Higgs 
configuration, it could well be that a scalar potential is generated 
due to infrared effects. Another possible modification
to the classical equations of motion is through higher derivative terms
in the effective action. The point is that, for example, a hypothetical term
$\sim (X^4)^2 (\del X^8)^4$ usually plays no role in low-energy considerations, but
with our background it would typically generate a mass-like term $\sim (X^4)^2$
and hence would be relevant.\footnote{
Quantum modifications of equations of motion and BPS conditions in $\cN=2$
SYM have been studied in \cite{rocek}.
}
The fact the single condition
\eq{xeq0} serves two different  purposes (supersymmetry and equations
of motion) is not obvious a priori.  This adds to our confidence that
it will eventually also arise on the SYM side.

Similar phenomena appear  in the cases of 1/8 supersymmetry with
non-vanishing field strength.  We recall that with the gauge field
turned on, on the YM side,  one can preserve 1/8 supersymmetry with a
self-dual as well as an anti self-dual configuration, while on the
D3-brane side, we saw that the configuration can preserve 1/8
supersymmetry only with a self-dual  but not an anti self-dual
Born-Infeld field.  The situation could be brought back into harmony
if we remember (again) that the analysis we did on the SYM side is a
classical one. We expect  that \cite{progress} if one does a careful
job of quantization, then the anti self-dual gauge theory
configuration can be seen to break all supersymmetries. A similar
situation occurs in the $\cN=2$  Seiberg-Witten theory where only
instantons (no anti-instantons) contribute to the prepotential.

For the case of our brane configuration 2 of a single D3-brane
preserving 1/4 of the supersymmetries, an intuitive way to think about
the pair of conditions \eq{two} is that before one takes the large $N$
limit, one has in fact a D3-brane with worldvolume in the 0389
directions intersecting a large stack of D3-branes with worldvolume in
the 0123 directions and sitting at $X^m=0$.  It would be interesting
to see how the Cauchy-Riemann 
condition \eq{holo} emerges \cite{progress} when one
takes the near horizon limit of the sugra solution for this system of
branes.

In this paper, we have treated the D3-brane as a brane probe in the
\ads background. This is appropriate for the large $N$ limit of the
AdS/gauge theory correspondence. However, as an independent question,
it would be interesting to work out the sugra solution with the
D3-brane as a brane source corresponding to the various cases we
considered here.  The metric should be asymptotically \ads and the
RR-flux should jump as one crosses the brane \cite{witten2}. It is
particularly interesting to obtain the  sugra solution corresponding
to  the ``holomorphically" embedded case.

The agreement of supersymmetry is of course a necessary condition for
the desired duality to work. Having obtained $\cN=2,1$ four
dimensional gauge theories, one may wonder what can one learn about
their correlation functions from studying the corresponding sugra
picture  or vice versa. Many 3-point and 4-point  correlations
function of the $\cN=4$ superconformal Yang-Mills  has been computed
\cite{f1,liu,f2,f3,f4} and precise  agreements were found with the
direct gauge theory calculations.

%\bigskip
%\vspace{.1cm}
\newpage

\noindent{\large \bf Acknowledgments}

We would like to thank C. Bachas, L. Bonora,  J.-P. Derendinger,
P.M. Ho, E. Sezgin and Y.Y. Wu for discussions.  Support by the Swiss National
Science Foundation is gratefully acknowledged.

\ed